**Prediction model of band-gap for AX binary compounds by combination of density functional theory calculations and machine learning techniques**

Joohwi Lee[1,*], Atsuto Seko[1,2,3], Kazuki Shitara[1,2,4], Isao Tanaka[1,2,3,4]

[1]*Department of Materials Science and Engineering, Kyoto University, Kyoto, 606-8501, Japan*

[2]*Elements Strategy Initiative for Structure Materials (ESISM), Kyoto University, Kyoto 606-8501, Japan*

[3]*Center for Materials Research by Information Integration, National Institute for Materials Science (NIMS), Tsukuba 305-0047, Japan*

[4]*Nanostructures Research Laboratory, Japan Fine Ceramics Center, Nagoya 456-8587, Japan*

[*]Corresponding author. Tel.: +81 75 753 5435

E-mail address: lee.joohwi@gmail.com (Joohwi Lee)

**Abstract**

Machine learning techniques are applied to make prediction models of the $G_0W_0$ band-gaps for 156 AX binary compounds using Kohn-Sham band-gaps and other fundamental information of constituent elements and crystal structure as predictors. Ordinary least square regression (OLSR), least absolute shrinkage and selection operator (LASSO) and non-linear support vector regression (SVR) methods are applied with several levels of predictor sets. When the Kohn-Sham band-gap by GGA (PBE) or modified Becke-Johnson (mBJ) is used as a single predictor, OLSR model predicts the $G_0W_0$ band-gap of a randomly selected test data with the root mean square error (RMSE) of 0.54 eV. When Kohn-Sham band gap by PBE and mBJ methods are used together with a set of various forms of predictors representing constituent elements and crystal structures, RMSE decreases significantly. The best model by SVR yields the RMSE of 0.18 eV. A large set of band-gaps estimated in this way should be useful as predictors for materials exploration.

## I. Introduction

The band-gap is a simple but important parameter to characterize semiconductors and insulators for optical and electronic applications [1]. To explore a material with a desired property from a large set of compounds, it can be a good starting point to know the band-gap of all compounds in a library. However, experimental data for band-gaps are still limited. Accurate measurement of band-gap requires high quality single-crystals. However, they are often hard to be synthesized. By virtue of recent advances in computational power and techniques, first principles calculation becomes the common approach to estimate the band-gap of a large number of compounds. Actually, the band-gap has been already computed for over 40,000 compounds and they are included in several open-access databases [2-4]. Most of them were computed by the density functional theory (DFT) calculation [5] with the generalized gradient approximation (GGA) of Perdew-Burke-Ernzerhof (PBE) [6] exchange-correlation functional. The Kohn-Sham gap (KS-gap), namely the difference between lowest unoccupied and highest occupied eigenvalues is typically used as an approximated band-gap. However, there is a well-known drawback; the KS-gap by GGA underestimates the band-gap significantly. As an example, Fig. 1 shows the comparison of the KS-gap by GGA-PBE and experimental band gap for 32 AX compounds included in our dataset that will be shown later. The root-mean-square (RMS) difference between PBE and experimental band-gaps is as large as 2.25 eV.

A practical remedy for the underestimation is the use of a new exchange-correlation functional such as the modified Becke-Johnson (mBJ) functional by Tran and Blaha [7]. Although the mBJ functional has been successful for evaluating the band-gap in a number of compounds, there seems to be small but systematic inconsistency between mBJ KS-gap and experimental band-gap. As shown in Fig. 1, the deviation between the mBJ KS gap by and experimental band-gap is 0.73 eV for the 32 AX compounds as an average. Despite the improvement from PBE, researchers often need to evaluate the band-gap more accurately. There are at least two different approaches for the improvement. One is so called the delta self-consistent-field ($\Delta$SCF) method to evaluate the band-gap from the differences in total energies [8]. The other uses hybrid functional methods [9] or GW calculations based on the many-body perturbation theory [10]. The downside of the latter methods is their high computational costs. It is still difficult to establish a large database of band-gap by such methods. Therefore, an alternative way to accurately estimate the band-gap using computationally affordable methods is desired.

Machine learning methods have been used for a number of purposes in condensed matter physics and materials science. The data-driven approaches to estimate a physical quantity are useful especially when a target property cannot be directly computed by the GGA level calculations. A prediction model can be constructed by a regression method taking the GGA level dataset and other fundamental information of constituent elements and crystal structure as predictors. Such methods have been applied to estimate a wide range of material properties such as the melting temperature [11,12], the ionic conductivity [13,14], the phase stability [15,16], the potential energy surface [17-19], the atomization energy of molecules [20,21], and so on. On the prediction of the band-gap, a few applications of regression methods have been reported [22-24]. Setwayan *et al*. estimated a relationship between PBE KS-gap and experimental band-gap by ordinary least squares regression (OLSR) from a dataset composed of about 100 compounds established on the database AFLOWLIB [2] including both of direct and indirect band-gaps [22]. Dey *et al*. predicted the direct band-gap of about 200 ternary chalcopyrite compounds from 28 experimental band-gap observations by OLSR, sparse partial least square regression and least absolute shrinkage and selection operator (LASSO) methods with predictors such as valence, atomic number, melting point, electronegativity and pseudo-potential radii of each element [23]. Gu *et al*., applied support vector regression (SVR) and artificial neural network to predict experimental band-gaps of 25 binary and 31 ternary compounds with some element-specific predictors [24]. Despite these pioneering works, there is still a plenty of room for improvement by selecting better regression techniques and predictors for the machine learning. In this study, we construct prediction models of the band-gap using different kinds of predictors and regression techniques. Firstly, we estimate relationships between the band-gaps obtained by different levels of approximations in the first principles calculation. We set a target property to the quasi-particle gap (QP-gap) for AX binary compounds that is obtained by the $G_0W_0$ calculation with HSE06 functional instead of the experimental band-gap. As shown in Fig. 1, the $G_0W_0$ band-gap is averagely the closest to the experimental band-gap. Secondly, prediction models for the QP-gap by the $G_0W_0$ calculation, $E_g$ ($G_0W_0$), are constructed from DFT KS-gaps and fundamental information of constituent elements and crystal structures using OLSR, LASSO, and nonlinear SVR.

## II. Methodology

### A. Regression methods

Relationship between a target property and predictors can be obtained by regression methods. The reliability of the estimation depends on the quality of the training set, selection of predictors and regression method. For a given training set, it is essential to select “good” predictors and regression methods. In the present study, we compare OLSR, LASSO, and nonlinear SVR regression methods. In the OLSR regression, regression coefficients of predictors, $\boldsymbol{\beta}$, are determined from $n$ observations by minimizing the following minimization function $L(\boldsymbol{\beta})$, given by

$$L(\boldsymbol{\beta}) = \|\boldsymbol{y} - \boldsymbol{X}\boldsymbol{\beta}\|_2^2 \, , \tag{1}$$

where $\| \, \|_2$ denotes L2-norm. $\boldsymbol{X}$ and $\boldsymbol{y}$ are ($n \times p$) predictor matrix and $n$-dimensional vector of target property for training data, respectively. The LASSO regression enables us not only to provide a solution for linear regression but also to obtain a sparse representation with a small number of nonzero regression coefficients[25]. The LASSO minimization function is defined as

$$L(\boldsymbol{\beta}) = \|\boldsymbol{y} - \boldsymbol{X}\boldsymbol{\beta}\|_2^2 + \lambda \|\boldsymbol{\beta}\|_1 \, , \tag{2}$$

where $\| \, \|_1$ denotes the L1-norm. The parameter $\lambda$ controls the trade-off relationship between sparsity and accuracy. Meanwhile, nonlinear regressions are more flexible to express complex functions. The SVR [26] is a nonlinear regression method based on the kernel trick [27]. The SVR is a version of support vector machine for regression. The kernel function maps the original feature space of the considered data into a high-dimensional space, where the learning task is simplified. Here we used the SVR with Gaussian kernel function. The Gaussian kernel function between two data points $\boldsymbol{x}$ and $\boldsymbol{x}'$ with a parameter $\sigma^2$ is defined as

$$k(\boldsymbol{x}, \boldsymbol{x}') = \exp\left(-\frac{\|x - x'\|^2}{2\sigma^2}\right). \tag{3}$$

Internal parameters appeared in the SVR are optimized by minimizing the cross validation (CV) score. The SVR is performed using e1071 package [28] implemented in R [29].

### B. Computational detail of first principles calculation

We made a first-principles dataset composed of AX binary compounds of I-VII, II-VI, III-V and IV-IV elements. Cations include Li, Na, K, Rb, Mg, Ca, Sr, Ba, Zn, Cd, B, Al, Ga, In, Si and Ge. Anions include F, Cl, I, O, S, Se, Te, N, P, As, Sb, and C. In order to examine the construction of prediction models in a simple way, compounds with transition metals are excluded in the present study. For each compound, four kinds of the crystal structure were considered such as wurtzite- (WZ), cesium chloride- (CC), zincblende- (ZB) and rocksalt-type (RS) structures. Totally, we studied 216 combinations of elements and crystal structures. Firstly, we calculated $E_g$ (PBE) and $E_g$ (mBJ) for the 216 compounds. Among the 216 compounds, 156 compounds exhibiting positive $E_g$ (PBE) and $E_g$ (mBJ) were used for the regression. Then, $E_g$ was calculated by the $G_0W_0$ calculation based on one-electron states obtained by the DFT calculation with the HSE06 functional for the 156 compounds. In addition, the cohesive energy was calculated by the PBE functional to be used as a predictor for building prediction models of $E_g$ ($G_0W_0$).

All of first-principles calculations were performed by the project augmented wave (PAW) method [30,31] implemented in the Vienna *Ab-initio* Simulation Package (VASP) [32,33]. The cutoff energy was set to 500 eV for PBE and mBJ calculations. Before calculating band-gaps, the structure optimization was performed using PBE functional revised for solids (PBEsol) [34] until residual forces acting on atoms reach below 0.005 eV/Å. *k*-space sampling were performed using 8×8×6, 8×8×8, 8×8×8 and 12×12×12 Γ-centered meshes for WZ, RS and ZB, and CC, respectively. For the $G_0W_0$ calculation with HSE06 functional, a set of pseudo-potentials updated for GW calculations was used. The cutoff energy was set to 600 eV. The number of *k*-points was reduced to 4×4×3, 4×4×4, 4×4×4 and 6×6×6 for WZ, RS, ZB, and CC structures, respectively.

## III. Results and discussion

### A. Comparison of theoretical band-gaps

Among 216 AX binary compounds, we found experimental band-gaps of 32 compounds as shown in Appendix A. Theoretical and experimental band-gaps are compared in Fig. 1. The tendency of the band-gap accuracy with respect to methods is similar to previous reports [7,10]. Figure 2 shows the histogram of differences between the theoretical and experimental band-gaps. The mean value of the difference between PBE KS-gap and experimental gap is -2.11 eV and -1.38 eV for direct and indirect compounds, respectively. Their RMS differences are 2.45 eV and 1.77 eV, respectively. Using the mBJ functional, theoretical band-gaps are notably improved. However, a tendency to underestimate the band-gap can still be noted by the mBJ calculations. $G_0W_0$ calculations show much better description of the band-gaps not only for the mean value but also for the RMS difference. Since the number of experimental data is limited, we will use the $G_0W_0$ QS-gaps as the target property in the present study. Considering the fact that the experimental band-gaps by different groups are sometimes scattered especially for the indirect band-gaps, the present results may be applicable to estimate the real band-gaps.

### B. Correction of band-gaps of PBE and mBJ

Figure 3(a) shows the relationship between $E_g$ ($G_0W_0$) and $E_g$ (PBE) for 156 AX binary compounds. Both of direct and indirect band gap materials are shown together. A similar plot is made for $E_g$ ($G_0W_0$) versus $E_g$ (mBJ) in Fig. 3(b). Both of them exhibit almost the linear dependence. Prediction models of $E_g$ ($G_0W_0$) for AX binary compounds are then made by the OLSR only from $E_g$ (PBE) and $E_g$ (mBJ). To estimate the prediction error of the models, the whole dataset is randomly divided into a training data composed of three quarters of the whole dataset; the rest is used as a test data. The random selection of the training data set is repeated for 100 times. The prediction error is estimated as the RMS error (RMSE) for the test data and 10-fold CV score. We also evaluated the mean absolute percentage error (MAPE) for the test data defined as

$$\frac{1}{n_{\mathrm{test}}}\sum_{i=1}^{n_{\mathrm{test}}}\left|\frac{y_i-\hat{y}_i}{y_i}\right|\times 100 \quad (4)$$

where $y_i$, $\hat{y}_i$ and $n_{\mathrm{test}}$ denote the observed target property for data $i$, predicted target property for data $i$ and the number of test data, respectively. RMSE, CV score and MAPE are averaged over 100 trials. Table I summarizes RMSE, CV score and MAPE of the OLSR models constructed by $E_g$ (PBE) and $E_g$ (mBJ). The OLSR model with a single predictor, $E_g$ (PBE) or $E_g$ (mBJ) shows RMSE of 0.52 eV and 0.57 eV, respectively.

When both $E_g$ (PBE) and $E_g$ (mBJ) are used as predictors, the OLSR model shows RMSE of 0.36 eV, which is much smaller than those of the OLSR models with a single predictor. The same behavior can be seen in the CV score and MAPE. A linear model is obtained as $E_g$ ($G_0W_0$) = 0.82 $E_g$ (PBE) + 0.49 $E_g$ (mBJ) + 0.36 in the unit of eV. Physics behind the improvement by using two predictors is not clear at the moment. However, the correlation coefficient for the linear plot is increased when both of $E_g$(PBE) and $E_g$(mBJ) are used as predictors instead of a single KS gap, which is the phenomenological reason for the improvement. Typical result using the average of $E_g$ (PBE) and $E_g$ (mBJ) is shown in Fig. 3(c).

### C. Comparison of different prediction models

In order to further improve the prediction models for $E_g$ ($G_0W_0$), different kinds of predictors and regression techniques are compared. Predictors should be included in a library, or those easily made by combining the physical quantities in a library. Alternatively, the predictors may be routinely computed by affordable DFT calculations. In the present study, we firstly use quantities obtained by GGA-PBE calculations, i.e., the cohesive energy, $E_{coh}$, and crystalline volume per atom, $V$, in addition to $E_g$ (PBE) and $E_g$ (mBJ) as predictors. Other fundamental variables related to constituent elements are also taken as candidates of predictors. They are the formal ionic charge, $n$, period, $p$, in the periodic table, atomic number, $Z$, atomic mass, $m$, van der Waals radius, $r^{vdW}$, electronegativity, $\chi$, and first ionization energy, $I$ [35]. Since there are only four simple crystal structures examined in the present study, only the coordination number of cation, $C_N$, is used as a predictor to represent crystal structures. For a given binary compound, AX, the element-specific predictors are converted to symmetric forms [11] so that they become symmetric with respect to the exchange of atomic species in binary compounds. The symmetric forms include the sum ($\xi_A+\xi_X$), absolute value of the difference ($|\xi_A-\xi_X|$), and product ($\xi_A\xi_X$) forms, where $\xi_A$ and $\xi_X$ denote a predictor for cation A and anion X, respectively. Original and inverse forms of these predictors including those of symmetric forms are taken as candidates of predictors. Three kinds of predictor sets summarized in Table II are examined. The maximum number of predictors is 41 in the predictor set (3).

Using these predictor sets, prediction models of $E_g$ ($G_0W_0$) are constructed by three regression methods, i.e., OLSR, LASSO and SVR. It should be noted that the training data is standardized before performing the regressions. For OLSR, the stepwise optimization based on the Akaike information criterion (AIC) [36] is used to select an appropriate set of predictors. Fig. 4 and Table III show the prediction errors of OLSR, LASSO and SVR models. As described before, when both $E_g$ (PBE) and $E_g$ (mBJ) are used as predictors, the OLSR model provides the RMSE of 0.36 eV. The RMSE decreases with the increase of the number of predictors for OLSR. Among OLSR models, the OLSR model with the predictor set (3) gives the best RMSE of 0.21 eV. The results by LASSO are almost the same as those of OLSR with stepwise regression using the AIC. This is quite natural since the average numbers of selected predictors are almost the same in the two methods. When the same predictor set is used, the SVR model shows better prediction than the OLSR and LASSO model. The best RMSE using the SVR is 0.18 eV with the predictor set (3). The CV score and MAPE show the same tendency as the RMSE.

Figure 5 shows the comparison of $E_g$ directly calculated by $G_0W_0$ and predicted from the best SVR model. Results by one of 100 trials are displayed. Compared to the OLSR model only with $E_g$ (PBE) and $E_g$ (mBJ) (not shown), most of the data are located at much closer to the diagonal straight line, where the predicted $E_g$ is exactly the same as the observed one. As can also be understood from the improvement of RMSE, the prediction is improved by adding predictors on DFT properties and information of the constituent elements and structures. The errors of training and test data are similar and that the RMSE and CV score are very close, which implies that serious over-fitting problem does not occur.

Finally, we discuss the distribution of prediction error for 156 compounds by three regression methods. Figure 6 shows the histograms of the prediction error distribution for the OLSR, LASSO and SVR models with predictor set (3). To make the histogram, the RMSE was evaluated for 100 times. Two things can be learned by these plots. Firstly, the distribution of RMSE is different reflecting the difference between linear methods and non-linear kernel method, besides the smaller RMSE by the SVR model as an average. Secondly, there is no notable outlier among 156 compounds showing averaged RMSE of > 1 eV in all of three models. The results confirm the usefulness of the present models for establishing large database of band-gap with reasonable accuracy.

## IV. Conclusion

In this study, we have developed the prediction models of the $G_0W_0$ band-gaps for 156 AX binary compounds using KS band-gaps and other fundamental information of constituent elements and crystal structure as predictors. OLSR, LASSO and SVR methods were used with several levels of predictor sets (1) to (3). When $E_g$ (PBE) is used as a single predictor, OLSR model predicts the $G_0W_0$ band-gap of a randomly selected test data with RMSE of 0.52 eV. When $E_g$ (mBJ) is additionally used together with a set of various forms of predictors representing constituent elements and crystal structures, RMSE decreases significantly. The best model by SVR yields the RMSE of 0.18 eV, in which serious over-fitting problem does not take place. Once this type of well-corrected set of band-gaps is established, it should be useful as predictors for high throughput screening of materials.

## ACKNOWLEDGEMENTS

This work was supported by Grant-in-Aid for Scientific Research (A) and Grant-in-Aid for Scientific Research on Innovative Areas "Nano Informatics" (Grant No. 25106005) from the Japan Society for the Promotion of Science (JSPS), and Support program for starting up innovation hub on Materials research by Information Integration Initiative from Japan Science and Technology Agency. JL acknowledge Grant-in-Aid for International Research Fellow of JSPS (Grant No. 14F04376) and JSPS fellowships.

**Table I**. Prediction errors of OLSR models obtained with $E_g$ (PBE) and $E_g$ (mBJ) as predictors.

| Predictors | RMSE (eV) | CV score (eV) | MAPE (%) |
|---|---|---|---|
| $E_g$ (PBE) | 0.52 | 0.52 | 9.70 |
| $E_g$ (mBJ) | 0.57 | 0.57 | 10.78 |
| $E_g$ (PBE), $E_g$ (mBJ) | 0.36 | 0.37 | 7.31 |

**Table II**. Predictor sets used for regressions shown in Fig. 4.

| | Number of predictors | Predictors |
|---|---|---|
| Set (1) | 8 | $E_g$ (PBE), $E_g$ (mBJ),<br>Original and inverse forms of $E_{coh}$, $V$, $N_c$ |
| Set (2) | 28 | Set (1)<br>$\xi_A+\xi_X$, $\|\xi_A-\xi_X\|$, $\xi_A\xi_X$ forms of predictors on constituent elements |
| Set (3) | 41 | Set (2)<br>$1/(\xi_A+\xi_X)$, $1/\xi_A\xi_X$ forms of predictors on constituent elements |

**Table III.** Prediction errors for the OLSR, LASSO and SVR models.

| Regression methods | Predictor sets | Average number of selected predictors | RMSE (eV) | CV score (eV) | MAPE (%) |
|---|---|---|---|---|---|
| OLSR | Set (1) | 5.6 | 0.33 | 0.31 | 7.11 |
| | Set (2) | 19.8 | 0.25 | 0.24 | 6.63 |
| | Set (3) | 29.3 | 0.21 | 0.19 | 5.58 |
| LASSO | Set (1) | 6.6 | 0.32 | 0.32 | 6.98 |
| | Set (2) | 22.7 | 0.27 | 0.27 | 6.92 |
| | Set (3) | 31.2 | 0.21 | 0.22 | 5.38 |
| SVR | Set (1) | | 0.26 | 0.26 | 6.02 |
| | Set (2) | | 0.20 | 0.19 | 5.04 |
| | Set (3) | | 0.18 | 0.18 | 4.74 |

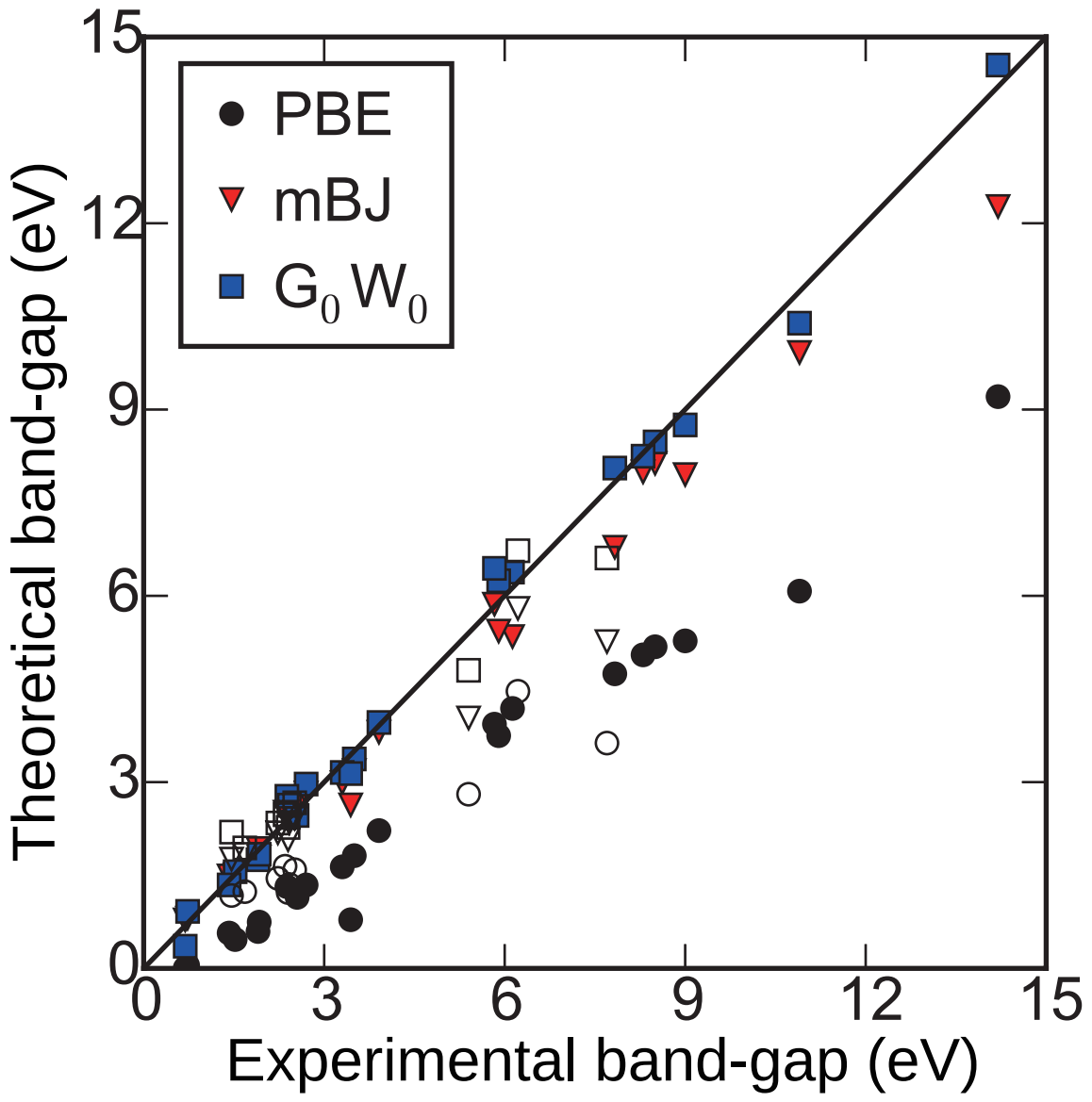

Fig. 1. Comparison of theoretical and experimental band-gaps of 32 AX compounds. The closed and open symbols indicate the direct (22 compounds) and indirect (10 compounds) band-gaps.

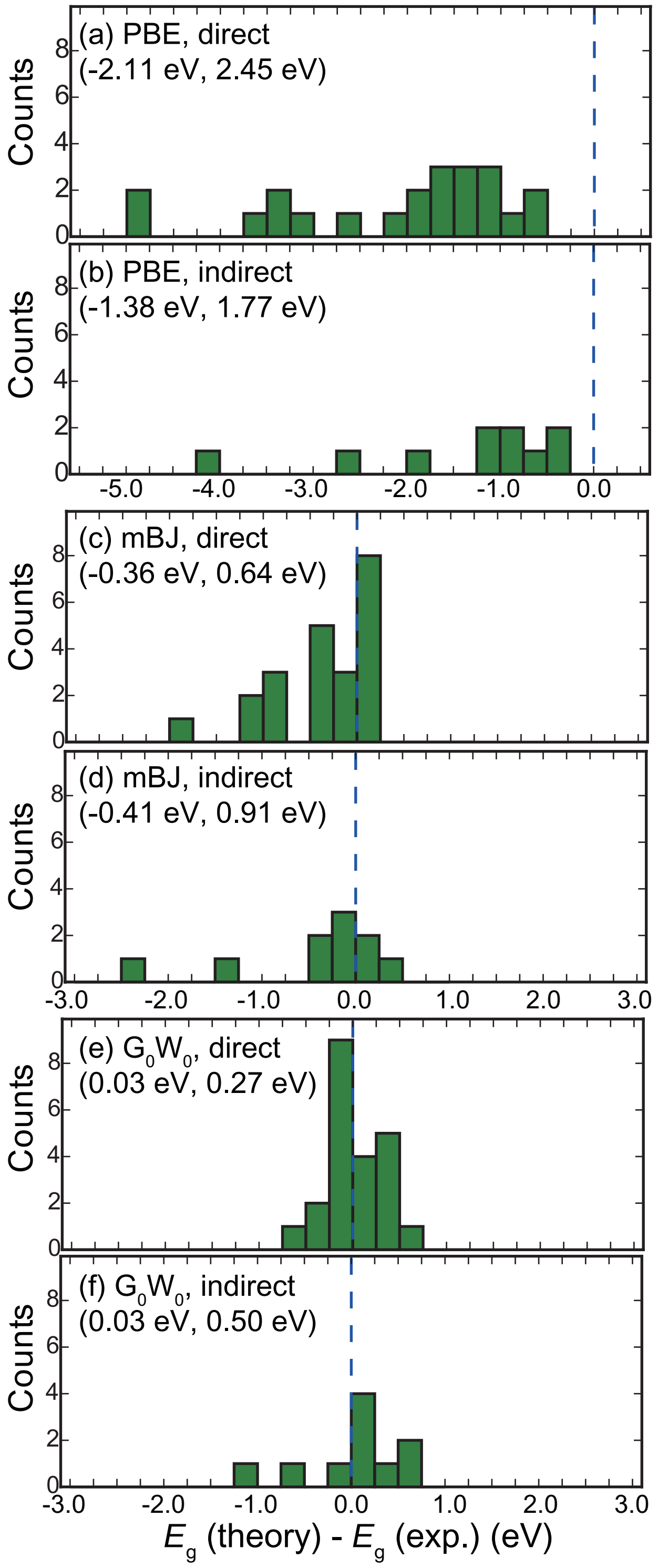

Fig. 2. Histogram of distribution of differences of theoretical and experimental band-gaps of 32 AX compounds. Theoretical band-gaps are obtained for direct-gap compounds by (a) PBE, (c) mBJ and (e) $G_0W_0$. Those for indirect-gap compounds by (b) PBE, (d) mBJ and (f) $G_0W_0$, respectively. Dashed vertical line corresponds to $E_g$ (theory) = $E_g$ (exp). Values in parentheses are mean and RMS differences.

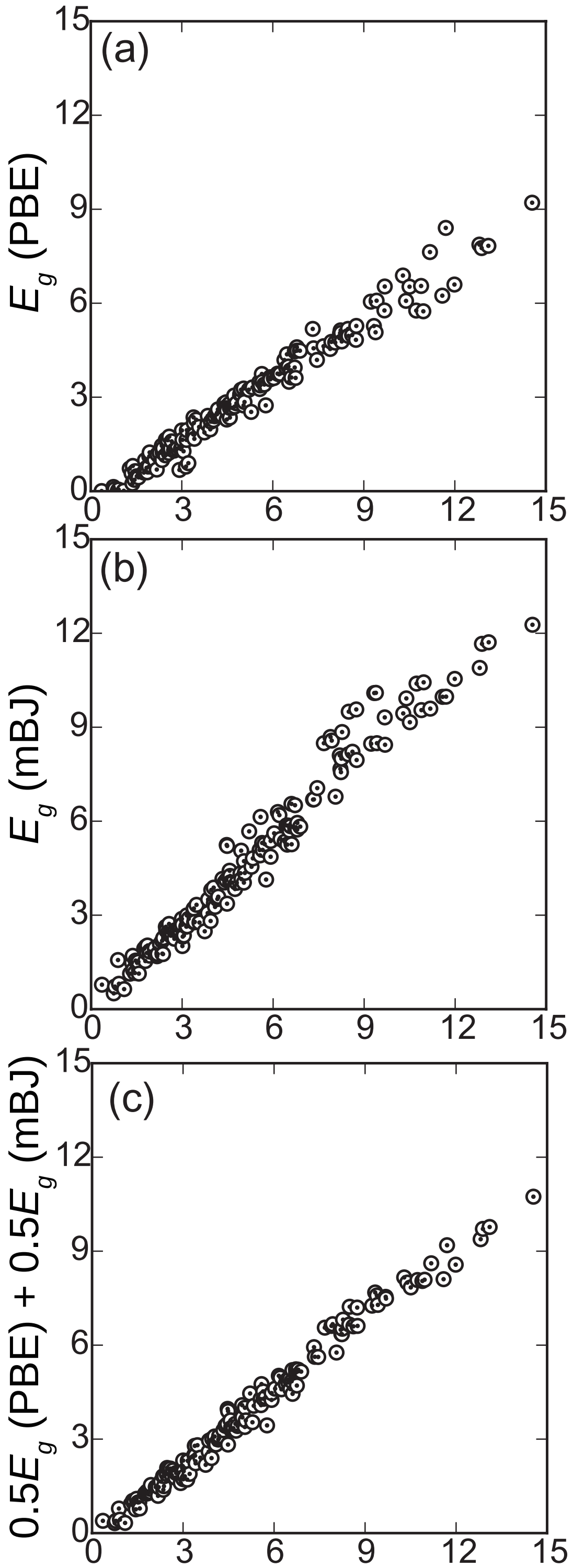

Fig. 3. Relationships between the QP-gap by the $G_0W_0$ calculation with HSE06 functional and KS-gap of (a) $E_g$ (PBE), (b) $E_g$ (mBJ), and (c) the average of $E_g$ (PBE) and $E_g$ (mBJ).

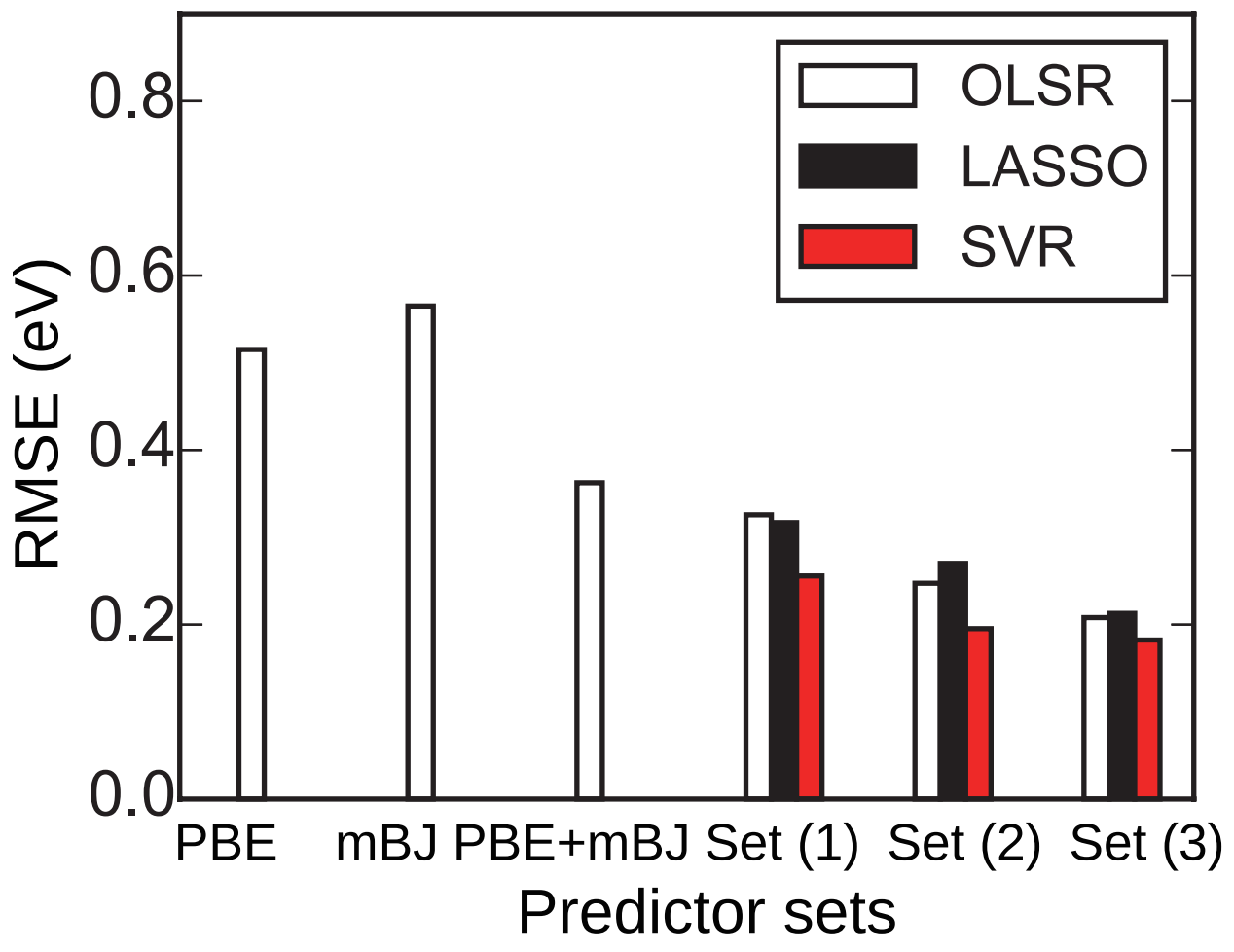

Fig. 4. Dependence of the prediction error on the predictor set. Black open bars and black and red closed bars denote the RMSEs by OLSR, LASSO and SVR models, respectively. The values are given in Tables I and III.

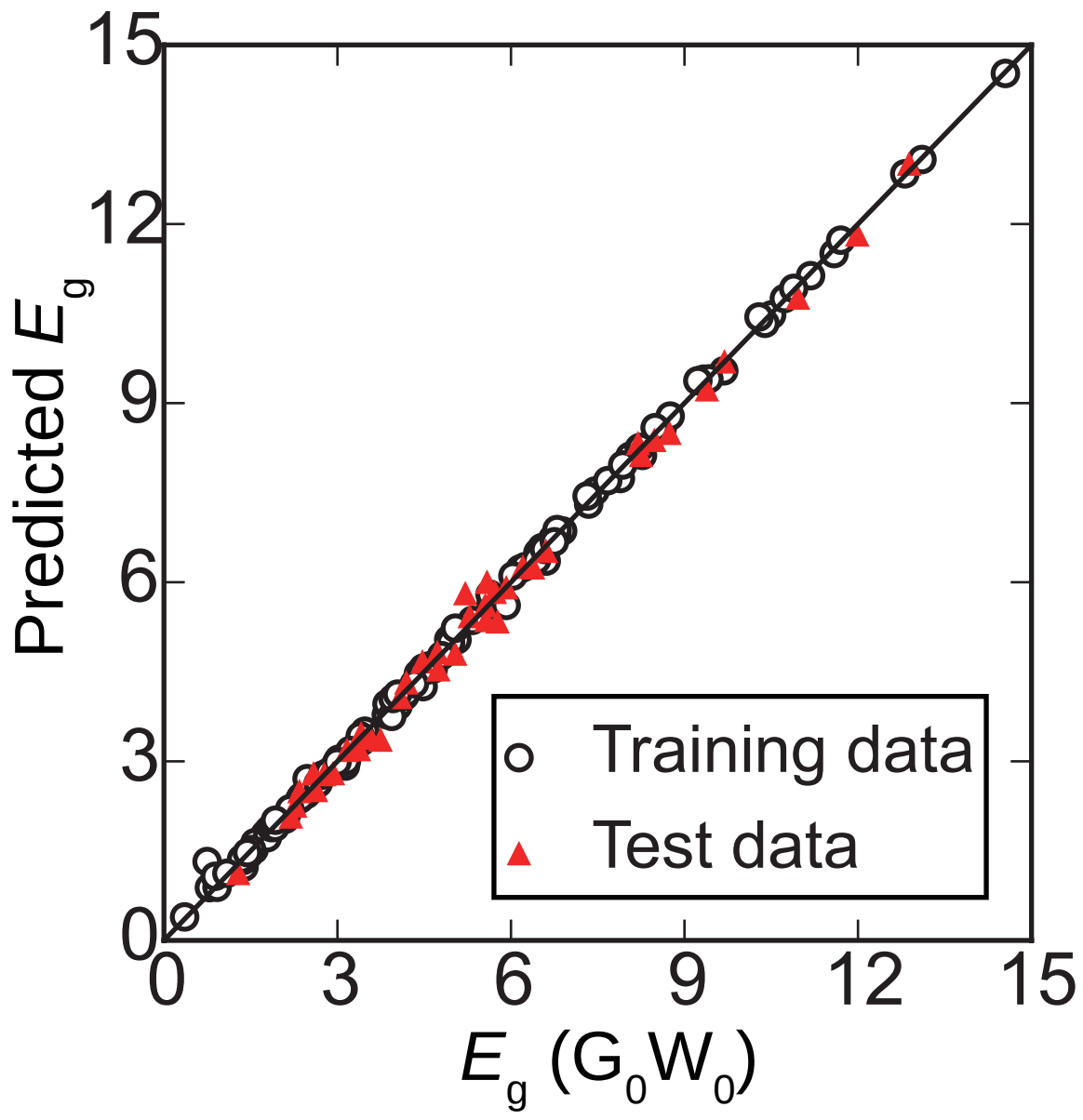

Fig. 5. (Color online) Comparison of $E_g$ calculated by $G_0W_0$ method and predicted with the SVR model using the predictor set (3). Results by one of 100 trials are shown here. Open circles and closed triangles show the training and test data, respectively.

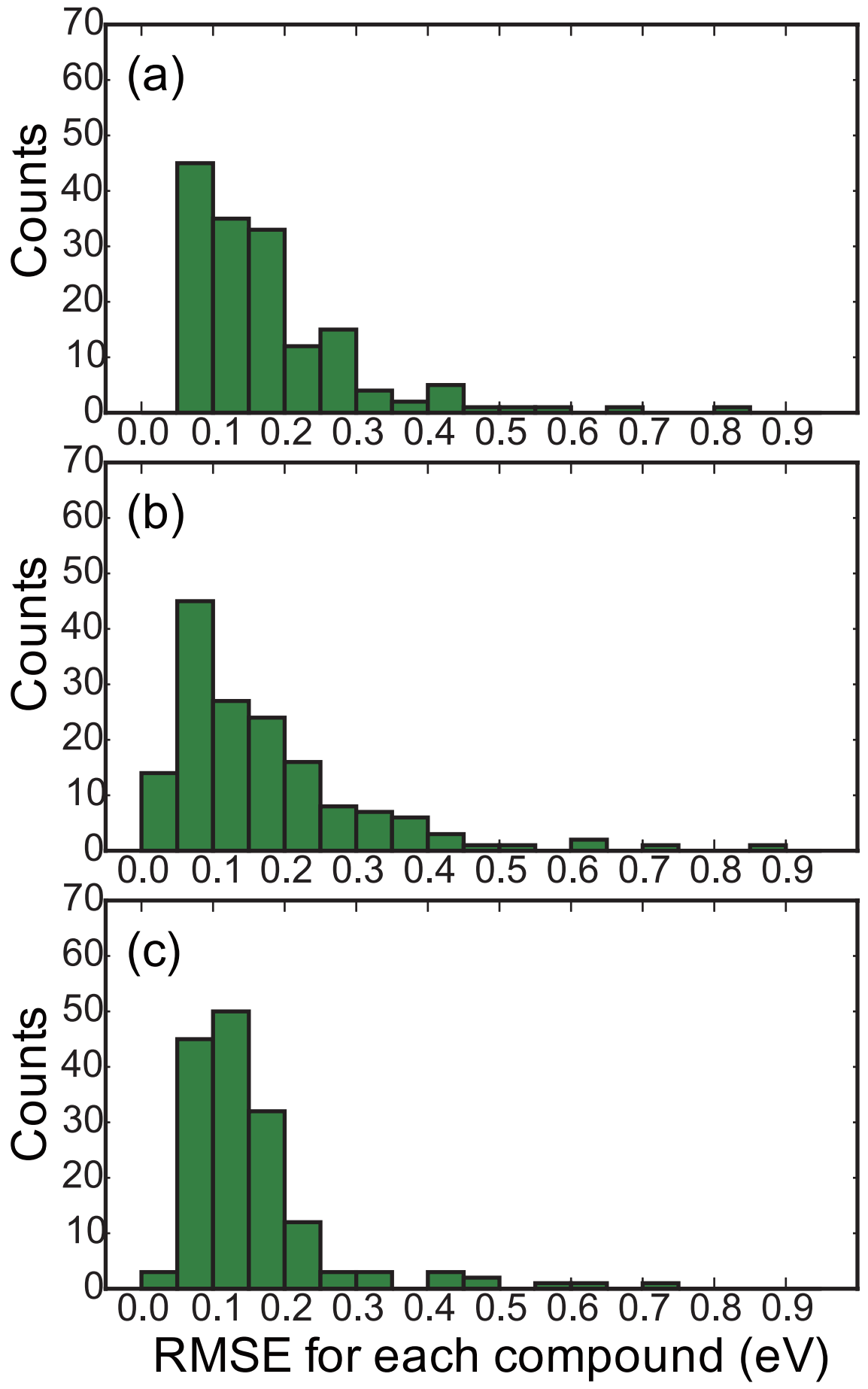

Fig. 6. (Color online) Histogram of distribution of RMSE for each compound using (a) OLSR, (b) LASSO and (c) SVR models with predictor set (3).

**Appendix A.** The band-gaps of the list of materials which were used at this study.

| Material | Crystal structure | $E_g$ (exp.) | Type[a] | $E_g$ (PBE) | $E_g$ (mBJ) | $E_g$ ($G_0W_0$, Target) |
|---|---|---|---|---|---|---|
| AlAs | RS | | M | | | |
| AlAs | CC | | M | | | |
| AlAs | ZB | 2.23 [c] | I | 1.45 | 2.18 | 2.34 |
| AlAs | WZ | | I | 1.74 | 2.38 | 2.59 |
| AlN | RS | | I | 4.59 | 5.73 | 6.79 |
| AlN | CC | | M | | | |
| AlN | ZB | | I | 3.31 | 4.81 | 5.33 |
| AlN | WZ | 6.13 [c] | D | 4.19 | 5.34 | 6.38 |
| AlP | RS | | M | | | |
| AlP | CC | | M | | | |
| AlP | ZB | 2.45 [c] | I | 1.59 | 2.42 | 2.67 |
| AlP | WZ | | I | 1.94 | 2.70 | 3.00 |
| AlSb | RS | | M | | | |
| AlSb | CC | | M | | | |
| AlSb | ZB | 1.68 [c] | I | 1.24 | 1.84 | 1.94 |
| AlSb | WZ | | D | 0.99 | 1.54 | 1.79 |
| BAs | RS | | M | | | |
| BAs | CC | | M | | | |
| BAs | ZB | 1.46 [c] | I | 1.17 | 1.76 | 2.20 |
| BAs | WZ | | I | 1.11 | 1.71 | 1.93 |
| BeO | RS | | I | 8.40 | 9.98 | 11.70 |
| BeO | CC | | I | 2.85 | 4.35 | 5.07 |
| BeO | ZB | | I | 6.88 | 9.44 | 10.29 |
| BeO | WZ | | D | 7.63 | 9.59 | 11.18 |
| BeS | RS | | I | 0.99 | 1.77 | 2.35 |
| BeS | CC | | M | | | |
| BeS | ZB | | I | 3.10 | 4.34 | 4.92 |
| BeS | WZ | | I | 3.64 | 4.86 | 5.92 |
| BeSe | RS | | I | 0.01 | 0.64 | 1.09 |
| BeSe | CC | | M | | | |
| BeSe | ZB | | I | 2.61 | 3.59 | 4.19 |
| BeSe | WZ | | I | 3.24 | 4.15 | 4.95 |
| BeTe | RS | | M | | | |
| BeTe | CC | | M | | | |
| BeTe | ZB | | I | 1.95 | 2.64 | 3.17 |
| BeTe | WZ | | I | 2.18 | 2.77 | 3.40 |
| BN | RS | | I | 1.24 | 1.76 | 2.37 |
| BN | CC | | M | | | |
| BN | ZB | 6.22 [c] | I | 4.46 | 5.80 | 6.73 |
| BN | WZ | | I | 5.18 | 6.69 | 7.32 |
| BP | RS | | M | | | |
| BP | CC | | M | | | |
| BP | ZB | 2.40 [c] | I | 1.22 | 2.07 | 2.27 |
| BP | WZ | | I | 0.99 | 1.94 | 2.10 |
| BSb | RS | | M | | | |
| BSb | CC | | M | | | |
| BSb | ZB | | I | 0.72 | 1.13 | 1.28 |
| BSb | WZ | | I | 0.81 | 1.25 | 1.38 |
| CaO | RS | 7.70 [d] | I | 3.63 | 5.26 | 6.61 |
| CaO | CC | | I | 2.74 | 4.14 | 5.77 |
| CaO | ZB | | I | 3.38 | 6.14 | 5.59 |
| CaO | WZ | | D | 3.24 | 5.68 | 5.21 |
| CaS | RS | | I | 2.29 | 3.37 | 4.48 |
| CaS | CC | | D | 1.27 | 2.35 | 3.06 |
| CaS | ZB | | I | 3.74 | 5.31 | 5.64 |
| CaS | WZ | | D | 2.70 | 3.84 | 4.74 |
| CaSe | RS | | I | 1.97 | 2.81 | 3.94 |
| CaSe | CC | | D | 0.69 | 1.69 | 2.17 |
| CaSe | ZB | | I | 3.28 | 4.72 | 5.04 |
| CaSe | WZ | | D | 2.39 | 3.26 | 4.09 |
| CaTe | RS | | I | 1.43 | 2.01 | 3.01 |
| CaTe | CC | | D | 0.13 | 0.50 | 0.75 |
| CaTe | ZB | | I | 3.05 | 3.93 | 4.73 |
| CaTe | WZ | | D | 1.87 | 2.49 | 3.75 |
| CdO | RS | | M | | | |
| CdO | CC | | M | | | |
| CdO | ZB | | M | | | |
| CdO | WZ | | D | 0.00 | 1.57 | 0.89 |
| CdS | RS | | I | 0.24 | 1.70 | 1.37 |
| CdS | CC | | M | | | |
| CdS | ZB | | D | 1.14 | 2.63 | 2.46 |
| CdS | WZ | 2.42 [c] | D | 1.22 | 2.72 | 2.57 |
| CdSe | RS | | M | | | |
| CdSe | CC | | M | | | |
| CdSe | ZB | 1.90 [c] | D | 0.59 | 1.93 | 1.75 |
| CdSe | WZ | | D | 0.65 | 2.00 | 1.81 |
| CdTe | RS | | M | | | |
| CdTe | CC | | M | | | |
| CdTe | ZB | 1.92 [c] | D | 0.75 | 1.79 | 1.84 |
| CdTe | WZ | | D | 0.79 | 1.84 | 1.92 |
| GaAs | RS | | M | | | |
| GaAs | CC | | M | | | |
| GaAs | ZB | 1.52 [c] | D | 0.46 | 1.51 | 1.56 |
| GaAs | WZ | | D | 0.48 | 1.48 | 1.52 |
| GaN | RS | | I | 0.60 | 2.03 | 1.86 |
| GaN | CC | | M | | | |
| GaN | ZB | 3.30 [c] | D | 1.64 | 2.99 | 3.16 |
| GaN | WZ | 3.50 [c] | D | 1.82 | 3.19 | 3.37 |
| GaP | RS | | M | | | |
| GaP | CC | | M | | | |
| GaP | ZB | 2.35 [c] | I | 1.65 | 2.52 | 2.48 |
| GaP | WZ | | D | 1.36 | 2.27 | 2.38 |
| GaSb | RS | | M | | | |
| GaSb | CC | | M | | | |
| GaSb | ZB | 0.73 [c] | D | 0.05 | 0.81 | 0.92 |
| GaSb | WZ | | D | 0.06 | 0.72 | 0.79 |
| GeC | RS | | M | | | |
| GeC | CC | | M | | | |
| GeC | ZB | | I | 1.63 | 2.44 | 2.54 |
| GeC | WZ | | I | 2.36 | 3.21 | 3.39 |
| InAs | RS | | M | | | |
| InAs | CC | | M | | | |
| InAs | ZB | | M | | | |
| InAs | WZ | | M | | | |
| InN | RS | | M | | | |
| InN | CC | | M | | | |
| InN | ZB | | M | | | |
| InN | WZ | 0.69 [c] | D | 0.00 | 0.78 | 0.36 |
| InP | RS | | M | | | |
| InP | CC | | M | | | |
| InP | ZB | 1.42 [c] | D | 0.57 | 1.49 | 1.34 |
| InP | WZ | | D | 0.64 | 1.55 | 1.47 |
| InSb | RS | | M | | | |
| InSb | CC | | M | | | |
| InSb | ZB | | M | | | |
| InSb | WZ | | M | | | |
| KCl | RS | 8.50 [d] | D | 5.18 | 8.13 | 8.48 |
| KCl | CC | | I | 5.05 | 7.67 | 8.22 |
| KCl | ZB | | I | 4.78 | 8.85 | 8.27 |
| KCl | WZ | | D | 4.77 | 8.57 | 7.93 |
| KF | RS | 10.90 [d] | D | 6.08 | 9.92 | 10.39 |
| KF | CC | | I | 6.55 | 9.55 | 10.89 |
| KF | ZB | | I | 5.07 | 10.10 | 9.39 |
| KF | WZ | | D | 5.27 | 10.09 | 9.33 |
| KI | RS | | D | 3.98 | 5.86 | 6.51 |
| KI | CC | | I | 3.40 | 5.27 | 5.72 |
| KI | ZB | | D | 3.95 | 6.51 | 6.72 |
| KI | WZ | | D | 3.73 | 6.19 | 6.21 |
| LiCl | RS | | D | 6.53 | 8.44 | 9.69 |
| LiCl | CC | | I | 4.56 | 6.70 | 7.34 |
| LiCl | ZB | | D | 6.08 | 8.49 | 9.42 |
| LiCl | WZ | | D | 6.05 | 8.47 | 9.23 |
| LiF | RS | 14.20 [b] | D | 9.21 | 12.27 | 14.55 |
| LiF | CC | | I | 7.87 | 10.89 | 12.81 |
| LiF | ZB | | D | 7.83 | 11.71 | 13.11 |
| LiF | WZ | | D | 7.76 | 11.66 | 12.88 |
| LiI | RS | | I | 4.37 | 5.25 | 6.47 |
| LiI | CC | | D | 2.27 | 3.39 | 4.05 |
| LiI | ZB | | D | 4.48 | 5.83 | 6.89 |
| LiI | WZ | | D | 4.49 | 5.95 | 6.80 |
| MgO | RS | 7.83 [b] | D | 4.74 | 6.78 | 8.06 |
| MgO | CC | | I | 2.53 | 4.55 | 5.29 |
| MgO | ZB | | D | 3.61 | 5.81 | 6.75 |
| MgO | WZ | | D | 3.49 | 5.70 | 6.54 |
| MgS | RS | 5.40 [c] | I | 2.80 | 4.03 | 4.80 |
| MgS | CC | | M | | | |
| MgS | ZB | | D | 3.49 | 5.08 | 5.65 |
| MgS | WZ | | D | 3.48 | 5.08 | 5.55 |
| MgSe | RS | | I | 1.80 | 2.79 | 3.41 |
| MgSe | CC | | M | | | |
| MgSe | ZB | | D | 2.66 | 4.04 | 4.66 |
| MgSe | WZ | | D | 2.66 | 4.05 | 4.58 |
| MgTe | RS | | I | 0.44 | 1.13 | 1.57 |
| MgTe | CC | | M | | | |
| MgTe | ZB | | D | 2.46 | 3.52 | 4.16 |
| MgTe | WZ | | D | 2.50 | 3.56 | 4.19 |
| NaCl | RS | 9.00 [d] | D | 5.27 | 7.95 | 8.75 |
| NaCl | CC | | I | 4.19 | 7.06 | 7.45 |
| NaCl | ZB | | D | 4.98 | 8.22 | 8.61 |
| NaCl | WZ | | D | 4.88 | 8.10 | 8.20 |
| NaF | RS | | D | 6.60 | 10.54 | 11.99 |
| NaF | CC | | I | 6.24 | 9.97 | 11.59 |
| NaF | ZB | | D | 5.74 | 10.43 | 10.96 |
| NaF | WZ | | D | 5.77 | 10.39 | 10.73 |
| NaI | RS | 5.90 [d] | D | 3.75 | 5.43 | 6.24 |
| NaI | CC | | D | 2.35 | 4.25 | 4.58 |
| NaI | ZB | | D | 3.86 | 5.84 | 6.58 |
| NaI | WZ | | D | 3.61 | 5.61 | 6.03 |
| RbCl | RS | 8.30 [d] | D | 5.05 | 8.00 | 8.25 |
| RbCl | CC | | I | 5.13 | 7.56 | 8.24 |
| RbCl | ZB | | I | 4.54 | 8.69 | 7.89 |
| RbCl | WZ | | D | 4.63 | 8.49 | 7.67 |
| RbF | RS | | I | 5.77 | 9.31 | 9.68 |
| RbF | CC | | D | 6.53 | 9.16 | 10.51 |
| RbF | ZB | | I | 4.83 | 9.57 | 8.74 |
| RbF | WZ | | D | 4.96 | 9.50 | 8.49 |
| RbI | RS | 5.83 [d] | D | 3.93 | 5.87 | 6.44 |
| RbI | CC | | I | 3.57 | 5.37 | 5.92 |
| RbI | ZB | | I | 3.85 | 6.55 | 6.60 |
| RbI | WZ | | D | 3.75 | 6.29 | 6.16 |
| SiC | RS | | M | | | |
| SiC | CC | | M | | | |
| SiC | ZB | 2.42 [c] | I | 1.35 | 2.35 | 2.63 |
| SiC | WZ | | I | 2.28 | 3.33 | 3.47 |
| SrO | RS | | I | 3.26 | 4.91 | 5.57 |
| SrO | CC | | I | 2.74 | 4.04 | 5.04 |
| SrO | ZB | | I | 2.58 | 5.21 | 4.48 |
| SrO | WZ | | D | 2.69 | 5.24 | 4.47 |
| SrS | RS | | I | 2.41 | 4.16 | 4.33 |
| SrS | CC | | I | 1.67 | 2.78 | 3.41 |
| SrS | ZB | | I | 3.11 | 5.06 | 4.95 |
| SrS | WZ | | D | 2.80 | 4.04 | 4.40 |
| SrSe | RS | | I | 2.13 | 3.07 | 3.84 |
| SrSe | CC | | I | 1.29 | 2.25 | 2.74 |
| SrSe | ZB | | I | 2.78 | 4.42 | 4.57 |
| SrSe | WZ | | D | 2.40 | 3.50 | 3.86 |
| SrTe | RS | | I | 1.64 | 2.32 | 3.02 |
| SrTe | CC | | I | 0.36 | 1.15 | 1.44 |
| SrTe | ZB | | I | 2.84 | 4.12 | 4.46 |
| SrTe | WZ | | D | 2.05 | 2.76 | 3.57 |
| ZnO | RS | | I | 0.89 | 2.89 | 3.22 |
| ZnO | CC | | M | | | |
| ZnO | ZB | | D | 0.68 | 2.52 | 2.92 |
| ZnO | WZ | 3.44 [b] | D | 0.78 | 2.63 | 3.14 |
| ZnS | RS | | M | | | |
| ZnS | CC | | M | | | |
| ZnS | ZB | 3.66 [b] | D | 2.22 | 3.80 | 3.96 |
| ZnS | WZ | | D | 2.29 | 3.88 | 4.04 |
| ZnSe | RS | | M | | | |
| ZnSe | CC | | M | | | |
| ZnSe | ZB | 2.71 [c] | D | 1.34 | 2.83 | 2.97 |
| ZnSe | WZ | | D | 1.39 | 2.88 | 2.99 |
| ZnTe | RS | | M | | | |
| ZnTe | CC | | M | | | |
| ZnTe | ZB | 2.38 [c] | D | 1.32 | 2.52 | 2.77 |
| ZnTe | WZ | | D | 1.36 | 2.50 | 2.85 |

[a] D, L, M denote direct band-gap, indirect band-gap, and metallic, respectively. Metallic materials were confirmed not by GW, but PBE calculations for saving time. Only positive band-gaps were used for regression.

[b] From ref.[37] [c] From ref.[38] [d] From ref. [39]